\documentclass[prl,aps,preprint,amsmath,showkeys,superscriptaddress]{revtex4-1}
\usepackage{graphicx}
\usepackage{dcolumn}
\usepackage{bm}
\usepackage{epstopdf}
\usepackage{url}
\begin{document}

\title{ Free-form intelligent hydrodynamic metamaterials enabled by extreme anisotropy }

\author{Fubao Yang}\thanks{These authors contributed equally to this work.}
\affiliation{Graduate School of China Academy of Engineering Physics, Beijing 100193, China}

\author{Yuhong Zhou}\thanks{These authors contributed equally to this work.}
\affiliation{Department of Physics, State Key Laboratory of Surface Physics, and Key Laboratory of Micro and Nano Photonic Structures (MOE), Fudan University, Shanghai 200438, China}

\author{Peng Jin}
\affiliation{Department of Physics, State Key Laboratory of Surface Physics, and Key Laboratory of Micro and Nano Photonic Structures (MOE), Fudan University, Shanghai 200438, China}

\author{Gaole Dai}\email{gldai@ntu.edu.cn}
\affiliation{School of Physical Science and Technology, Nantong University, Nantong 226019, China}

\author{Liujun Xu}\email{ljxu@gscaep.ac.cn}
\affiliation{Graduate School of China Academy of Engineering Physics, Beijing 100193, China}

\author{Jiping Huang}\email{jphuang@fudan.edu.cn}
\affiliation{Department of Physics, State Key Laboratory of Surface Physics, and Key Laboratory of Micro and Nano Photonic Structures (MOE), Fudan University, Shanghai 200438, China}

\date{\today}

\begin{abstract}

Intelligent metamaterials have attracted widespread research interest due to their self-adaptive capabilities and controllability. They hold great potential for advancing fluid control by providing responsive and flexible solutions. However, current designs of passive hydrodynamic metamaterials are limited by their fixed shapes and specific environments, lacking environmental adaptability. These two constraints hinder the broader application of hydrodynamic metamaterials. In this work, we propose a design for passive intelligent metashells that utilize extremely anisotropic parameters to endow hydrodynamic metamaterials with self-adaptive abilities and free-form shapes. Achieving the required anisotropic parameters is challenging, but we ingeniously accomplished this by creating isobaric conditions through increasing the water height in the shell region. We validated the design through finite-element simulations. This approach overcomes the limitations of existing passive hydrodynamic metamaterials, enhancing their intelligent behavior. Our model improves the flexibility and robustness of hydrodynamic metamaterials in complex and dynamic environments, providing insights for future designs and practical applications.

\end{abstract}


\maketitle

\section{Introduction}

Metamaterials are artificial materials that exhibit physical properties not found in natural materials~\cite{ZheNM12,WegenerScience13,KadicRPP13,YangRMP24}, representing a significant technological breakthrough impacting humanity. In recent years, thanks to the development of metamaterials, hydrodynamic metamaterials have rapidly advanced~\cite{UrzhumovPRL11,ParkPRL19,BoRPL21,TayNSR21,ChenInn23,DaiPRE23,ChenPOF24,ChenPNAS22,JiangAM24,ParkEML21,ParkPRAp19,LiPRL18,ZouPRL19,HanPRL22,PangPOF22,ChenDro23,WuPOF24,PangPOF24}. By appropriately simplifying the Navier-Stokes equations, the Darcy's law and Hele-Shaw model can be used to describe creeping or Stokes flow~\cite{ChenInn23,DaiPRE23,ChenPOF24,ChenPNAS22,JiangAM24}. On this basis, through the introduction of transformation theory~\cite{LeonhardtScience06,PendryScience06} and scattering cancellation technology~\cite{CummerPRL08,AluPRL09,HanPRL14}, many functional hydrodynamic metamaterials have emerged, such as the cloak~\cite{ChenInn23,DaiPRE23}, evener~\cite{ChenPOF24}, concentrator~\cite{ParkEML21}, rotator~\cite{ParkPRAp19}, and multifunctional switch~\cite{ChenPNAS22,JiangAM24}. These metamaterials not only enable precise control of fluids but also possess ``invisibility'' property, meaning they can manipulate fluids without disturbing the background physical fields, thus avoiding detection. The "invisibility" property makes hydrodynamic metamaterials particularly important in applications requiring high precision and stability, such as microfluidic technology and biomedical fields~\cite{EricNat14}. Fluid control is crucial in these areas, and the development of hydrodynamic metamaterials offers new paths for achieving more efficient and reliable solutions.

However, as the functional requirements become increasingly complex, the intelligent performance demands on hydrodynamic metamaterials continue to rise. Currently, existing passive hydrodynamic metamaterials, once designed, can only function in their pre-defined shapes within specific environments. Any changes in shape or background parameters often lead to the failure of these metamaterials, causing disturbances in the background physical fields. The fixed shape and working environment are two significant limitations that hinder their practical applications, necessitating urgent solutions.

\begin{figure}
	\centering
	\includegraphics[width=\linewidth]{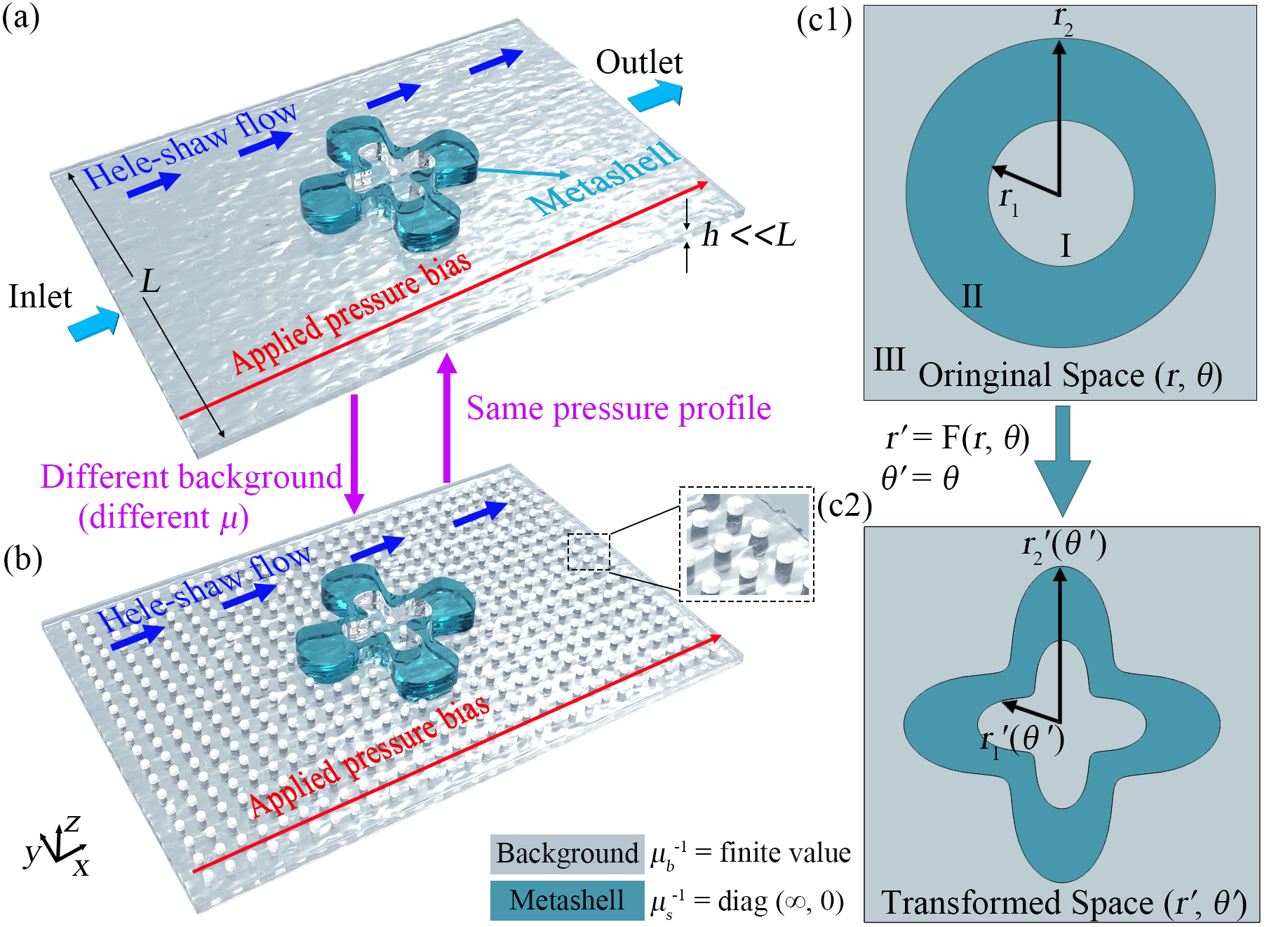}
	\caption{(a),(b) Illustrations of an intelligent hydrodynamic metashell in Hele-Shaw  cells with different background dynamic viscosities. The length and height of the cell is $L$ and $h$. Two boundaries parallel to the y-z plane are the inlet and outlet of the fluid. The other boundaries are non-slip. A pressure bias is applied along the x-axis. The metashell is free-form and filled with fluid. We add uniformly arranged pillars to the background in (b) to effectively change its dynamic viscosity. The dark blue arrows show the flow direction. (c1),(c2) Diagram of a radial coordinate transformation to transform the circular shell (c1) into the irregular shaped shell (c2). Background dynamic viscosity has a finite value, and the metashell dynamic viscosity satisfies $\mu_s^{-1} = \text{diag}(\infty,0)$. The inner and outer radii in the original space are $r_1$ and $r_2$, and $r_{1}'(\theta')$ and $r_{2}'(\theta')$ in the transformed space.}
	\label{f1}
\end{figure}

To address these two challenges, we propose a design for intelligent hydrodynamic metamaterials based on shells with extremely anisotropic parameters. We consider a model within Hele-Shaw cell~\cite{Panton13}. It describes the Stokes flow between two parallel plates, where the distance separating them is much smaller than their dimensions. Consequently, the fluid exhibits characteristics of creeping flow, which can be described using the Hele-Shaw model. Figs.~\ref{f1}(a) and \ref{f1}(b) are schematic diagrams illustrating the performance of intelligent hydrodynamic metashell within Hele-Shaw cells under different dynamic viscosity backgrounds. We use a four-leaf shape to represent that the metashell is free-form. Compared to Fig.~\ref{f1}(a), Fig.~\ref{f1}(b) has a background with uniformly arranged pillars, which effectively changes the dynamic viscosity, making it different from Compared to Fig.~\ref{f1}(a). When the metashell, with its extremely anisotropic dynamic viscosity, is placed in the two different backgrounds shown in Figs.~\ref{f1}(a) and \ref{f1}(b), the external physical fields (pressure field and velocity field) in both backgrounds will remain undisturbed, identical to the pure background physical fields. Due to the extremely anisotropic of the metashell's parameter, it functions as a fluid concentrator, increasing the pressure gradient and velocity in the core region. Given the same pressure boundary conditions for the metashell in different backgrounds, the pressure distributions in the external region of the metashell will be the same. This means that the designed metashell can adapt to changing environments and permit a free-form shape, thus overcoming the aforementioned limitations of existing passive hydrodynamic metamaterials and exhibiting intelligence. Notably, achieving the condition of extremely anisotropic dynamic viscosity is a challenge. In this work, we have equivalently realized this by raising the fluid height in the metashell region, and the validity of our design has been confirmed through simulations. The proposed passive intelligent hydrodynamic metamaterials allows for flexibility and robustness in practical applications, ensuring reliable performance even in complex and dynamic environment.

\section{Theoretical Analysis}
We consider an irrotational potential flow within a Hele-Shaw cell, as shown in Fig.~\ref{f1}. The width and height of the cell are $L$ and $h$, respectively. A pressure bias is applied along the $x$ direction. We assume that the fluid is incompressible and demonstrates a creeping flow with small Reynold number, $Re\ll1$. If the cell is fulled occupied and $h\ll L$, the velocity $\bm{v}$ can be obtained under lubrication approximation~\cite{Panton13}. Thus the average velocity $\overline{\bm{v}}$ in the x-y plane can be written as the Hele-Shaw equation
\begin{equation}\label{E1}
	\overline{\bm{v}}=\frac{-h^2}{12\mu}\nabla P,
\end{equation}
where $\mu$ is the dynamic viscosity, and $\nabla P$ is the applied pressure bias. If we take $\bm{v}\approx\overline{\bm{v}}$, the 3D model can be regarded as a 2D one. Considering the law of continuity $\nabla\cdot\bm{v} = 0$, the Hele-Shaw equation can be written as  
\begin{equation}\label{E2}
	\nabla\cdot\left(\gamma\nabla P\right)=0,
\end{equation}
where $\gamma = h^2/12\mu$. Eq.~(\ref{E2}) has the same form as the Darcy's law. In this work, we regard $h$ as a constant and consider cylindrical coordinates. Then $\mu^{-1}$ is equivalent to the permeability of Darcy's law, which means we can design hydrodynamic metamaterials by modulating the inverse of dynamic viscosity in the same way that they modulate permeability in porous media. 

The first limitation mentioned above is the lack of intelligence. Inspired by chameleonlike metashells suggested in metamaterial design\cite{XuPRAp19-1,XuSC20,YangPRAp20,ZhangPRAp23}, we use materials with extremely anisotropic dynamic viscosity to bestow intelligence on metashells. See Fig.~\ref{f1}(c1). The whole system is divided into three regions. Region I (core) and III are the background with the same dynamic viscosity $\mu_{b}$. Region II (shell) is the metashell with extremely anisotropic dynamic viscosity $\bm{\mu_s}$, i.e., $\bm{\mu}_s^{-1} = {\rm{diag}}(\mu_r^{-1},\mu_\theta^{-1}) = {\rm{diag}}(\infty,0)$. For a Laplacian-type governing equation, scattering-cancellation technology is a powerful way to design the key parameters. Therefore, by solving Eq.~(\ref{E2}) directly [Supplemental Material Sec. I], we obtain the effective dynamic viscosity $\mu_{e}$ of core-shell structure (regions I and II) in Fig.~\ref{f1}(b), which can be expressed as
\begin{equation}\label{E3}
	\mu_{e}^{-1} = m\mu_{r}^{-1}\frac{\left(1+p^{m}\right)\mu_{b}^{-1} + \left(1-p^{m}\right)m\mu_{r}^{-1}}{\left(1-p^{m}\right)\mu_{b}^{-1} + \left(1+p^{m}\right)m\mu_{r}^{-1}},
\end{equation}
where $m = \sqrt{\mu_{\theta}^{-1}/\mu_{r}^{-1}}$ and $p = \left(r_{1}/r_{2}\right)^{2}$. $r_1$ and $r_2$ are the inner and outer radii of the shell, respectively. Substituting $\mu_r^{-1} = \infty$ and $\mu_\theta^{-1} = 0$ into Eq.~(\ref{E3}), we have
\begin{equation}\label{E4}
	\mu_{e}^{-1} = \mu_{b}^{-1},
\end{equation}
which means the dynamic viscosity of core-shell structure are the same as that of the background. Therefore, the metashell behaves chameleonlike property. No matter how the background changes, the metashell will adapt to the background dynamic viscosity, without disturbing the distribution of pressure field in region III. Because of the extremely anisotropic dynamic viscosity, the metashell behaves as a perfect concentrator, that is, we theoretically design an intelligent hydrodynamic concentrator with extremely anisotropic materials. 

We further prove that such metashell can be irregular shaped.
It is easy to check Eq.~(\ref{E2}) satisfies the transformation theory. Considering a original space ($r$,$\theta$) and a transformed space ($r'$,$\theta'$), the transformation rule is $\bm{\gamma}' = \textbf{J}\bm{\gamma}\textbf{J}^{\rm{T}}/\rm{det}\textbf{J}$. Here $\bm{\gamma}'$ and $\bm{\gamma}$ are the parameters in the transformed and original space, respectively. $\textbf{J}$ is the Jacobian matrix denoted by $\textbf{J} = \partial(r',\theta')/\partial(r,\theta)$, and $\textbf{J}^{\rm{T}}$ is the transpose of $\textbf{J}$. Regarding $h$ as a constant, the transformation rule can be expressed as
\begin{equation}\label{E5}
	{\bm{\mu}'}^{-1}=\frac{\textbf{J}\bm{\mu}^{-1}{\rm \textbf{J}^{T}}}{\det\textbf{J}}.
\end{equation}
We consider a radial coordinate transformation $r' = {\rm{F}}(r,\theta)$ and $\theta' = \theta$ to transform the circular shell [with $\bm{\mu}_s^{-1} = {\rm{diag}}(\infty,0)$] shown in Fig.~\ref{f1}(c1) into the irregular shaped shell in Fig.~\ref{f1}(c2). According to Eq.~(\ref{E5}), the transformed dynamic viscosity of the metashell ${\bm{\mu}_{s}'}^{-1}$ is derived as
\begin{equation}\label{E6}
	{\bm{\mu}_{s}'}^{-1}=\left[
	\begin{matrix}
		\mu_{11}^{-1} & \mu_{12}^{-1}\\
		\mu_{21}^{-1} & \mu_{22}^{-1}
	\end{matrix}
	\right],
\end{equation}
where $\mu_{11}^{-1} = \infty$, $\mu_{12}^{-1}$ = $\mu_{21}^{-1}$ are finite values depending on the transformation, and $\mu_{22}^{-1} = 0$.  The detailed theoretical derivation is provided in Sec. II in Supplemental Material. By substituting Eq.~(\ref{E6}) into Eq.~(\ref{E2}), the Laplace equation in cylindrical coordinates is written as 
\begin{eqnarray}\label{E7}	
	\frac{1}{r'}\frac{\partial P}{\partial r'} +\frac{\partial^{2}P}{\partial r'^{2}} +\frac{\mu_{12}^{-1}}{\mu_{11}^{-1}}\frac{2}{r'}\frac{\partial^{2}P}{\partial r'\partial\theta'}
	+\frac{\mu_{12}^{-1}}{\mu_{11}^{-1}}\frac{1}{r'}\frac{\partial \mu_{12}}{\partial\theta}\frac{\partial P}{\partial r} + \frac{\mu_{22}^{-1}}{\mu_{11}^{-1}}\frac{1}{r'^{2}}\frac{\partial^{2}P}{\partial\theta'^{2}} =0.
\end{eqnarray}
It is clear that the off-diagonal components of ${\bm{\mu}_{s}^{-1}}'$ have little effect on fluid transport and can be omitted. By setting them as zero, we finally obtain the simplified ${\bm{\mu}_{s}^{-1}}'$, that is,
\begin{equation}\label{E8}
	{\bm{\mu}_{s}'}^{-1} =  {\rm{diag}}(\infty,0).
\end{equation}
According to the the theoretical analysis above, we find that metashells with extremely anisotropic dynamic viscosity meet the requirements of intelligence and irregular shape. Moreover, we can get the same conclusion from the view of null medium~\cite{SunOE19,FakheriPRAp20,BaratiSedehPRAp20,ChenIJTS22}. The detailed derivation can be found in Supplemental Material Sec. III. In what follows, we perform finite-element simulations using COMSOL Multiphysics to confirm our theoretical predictions.

\section{Results and Discussion}
\begin{figure}
	\centering
	\includegraphics[width=\linewidth]{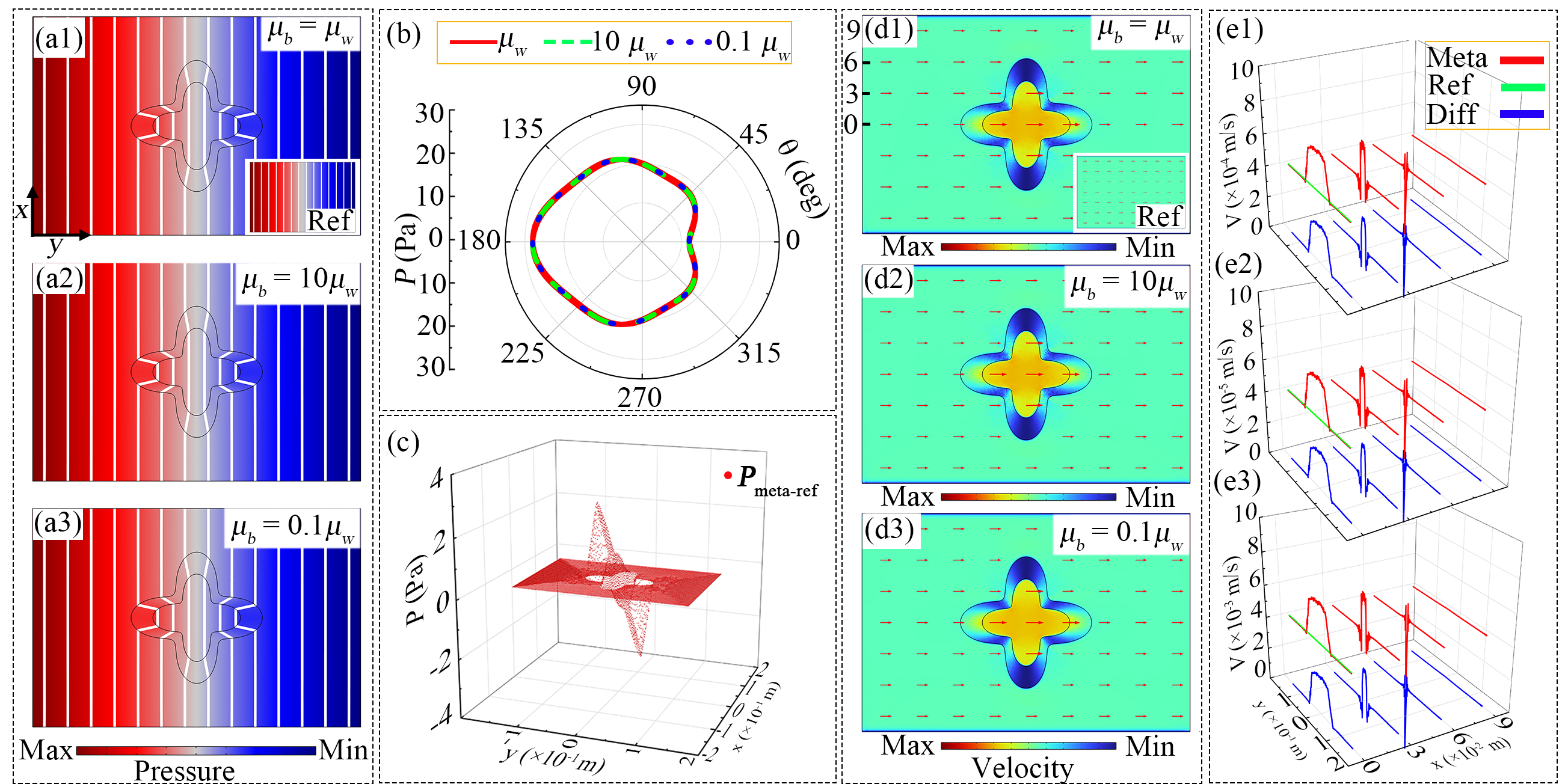}
	\caption{2D simulation results of the metashell. We choose four-leaf clover as the shape, which is described by $r_1(\theta) = \cos(4\theta)+3$ cm and $r_2(\theta) = 1.5[\cos(4\theta)+3]$ cm. The width and length of Hele-Shaw cell is 20 cm and 30 cm, respectively. (a1)-(a3) Pressure distributions under three different dynamic viscosity backgrounds. Maximum pressure is 40 pa, and minimum is 0 pa. (b) Pressure data charts of (a1)-(a3) along the outer contour of the metashell. The red solid line, green dashed line, and blue dotted line represent pressure data for $\mu_b = \mu_w$, $10\mu_w$, and $0.1\mu_w$, respectively.(c) The pressure difference between the metashell and the reference on the xy-plane when $\mu_b = \mu_w$. (d1)-(d3) Corresponding velocity distributions. The red arrows represent the normalized velocity, with the direction indicating the velocity direction and size representing the velocity magnitude. Maximum for (d1), (d2), and (d3) are 10$^{-3}$, 10$^{-4}$, and 10$^{-2}$~m~s$^{-1}$, respectively, with minimum being the same at 0~m~s$^{-1}$. (e1)-(e3) Velocity data charts along four lines at x = 0, 3, 6, and 9 cm [shown in (d1)]. The red lines represent the metashell. The green lines represent the reference. Since the reference has equal velocity everywhere, we only plot one line for simplicity and clarity. The blue lines at the bottom represent the velocity difference, equal to the velocity of metashell minus the velocity of reference.}
	\label{f2}
\end{figure}

Firstly, we perform 2D simulations and choose water with dynamic viscosity $\mu_w$ as the liquid material. We design four-leaf clover as the shape of the irregular shell, whose dynamic viscosity is diag($\infty$,0). To verify the metashell's intelligence, we put it in backgrounds with different dynamic viscosity, that is, $\mu_b = \mu_w$, 10$\mu_w$, and 0.1$\mu_w$, respectively. Figs.~\ref{f2}(a1)-(a3) depict the corresponding pressure fields. The metashell behaves as a perfect concentrator, making a larger pressure gradient in region I. We can qualitatively observe that this metashell can work in different backgrounds without disturbing the pressure distribution. For quantitive accuracy analysis, we read the pressure distribution along the outer contour of the shell under three backgrounds and draw it as Fig.~\ref{f2}(b). The three sets of data coincide, proving the accuracy of the theory. To quantitatively characterize the influence on the background, we use pure background as the reference group and subtract the pressure distribution of the reference group from the pressure distribution in Fig.~\ref{f2}(a1). The result is shown in Fig.~\ref{f2}(c). The red dots are the pressure differences, which is almost zero in region III. It means the metashell nearly has no effect on region III. 

We also show the corresponding velocity distributions in Figs.~\ref{f2}(d1)-(d3). The velocity in region III is uniform and undisturbed even though the background changes, consistent with theory. Since the metashell is a concentrator, the velocity is higher in region I. We further read their velocity data along four lines at x= 0,3,6,9 cm, and subtract the corresponding data of the reference group on four line segments to quantitatively demonstrate their influence on the background velocity field. See Fig.~\ref{f2}(e1). It depicts the velocity distribution when $\mu_b = \mu_w$. The blue lines at the bottom represent the differences at four positions, which are equal to 0 in region III, indicating that the metashell has no effect on the background. We can get the same conclusion from Fig.~\ref{f2}(e2) and \ref{f2}(e3). However, normal transformed metashell is limited to a fixed background and any variation in the background dynamic viscosity can lead to device failure (Supplemental Material Sec. IV). In addition, the intelligent metashell still works when the shape is asymmetrical, which proves the robustness of the theory (Supplemental Material Sec. V). 

\begin{figure}[ht!]
	\centering
	\includegraphics[width=\linewidth]{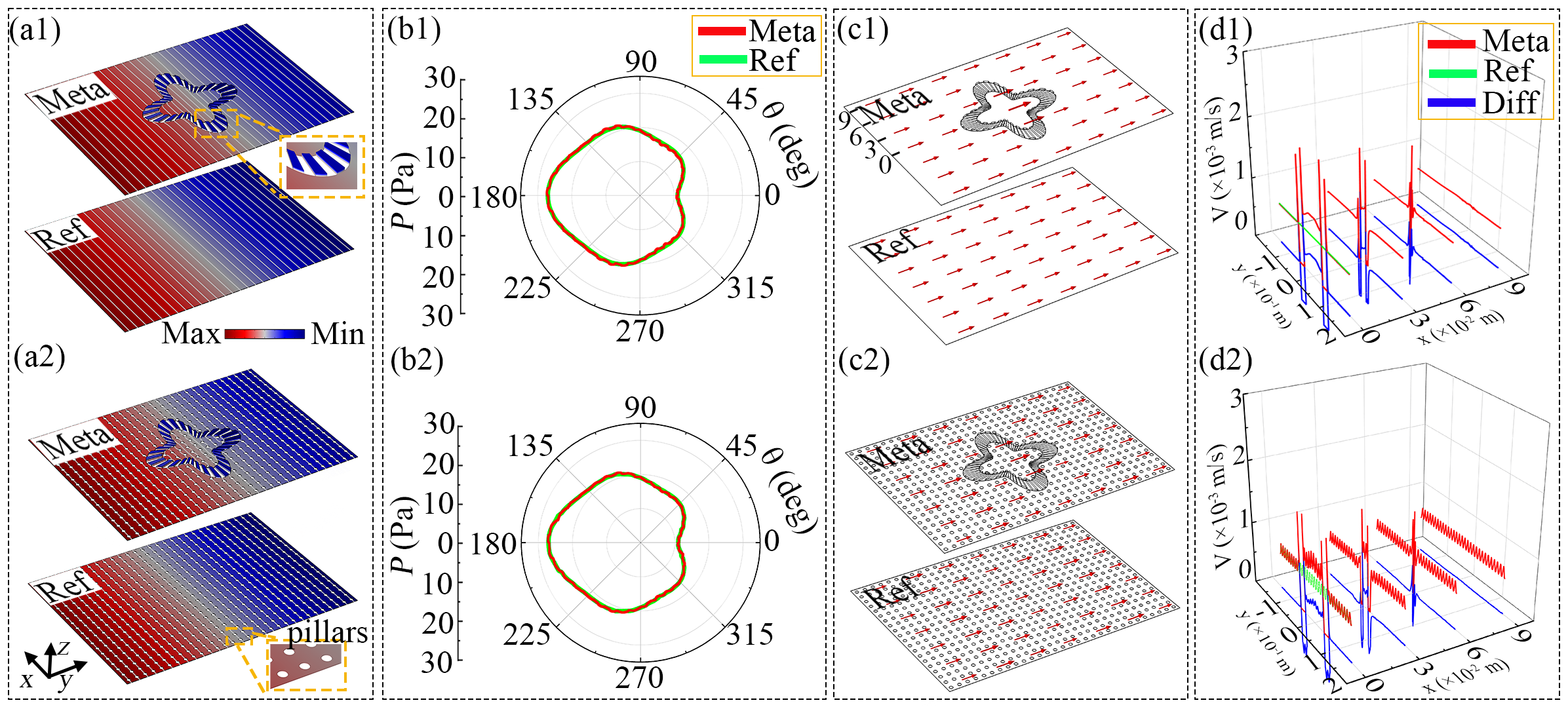}
	\caption{3D simulation results of the metashell. The size of Hele-Shaw cell is 20 cm $\times$ 30 cm $\times$ 0.02 cm. The height of the water in the metashell is 0.2 cm. The radius of pillars is 0.2 cm. (a1),(a2) Pressure distributions of the metashell and reference groups under two different dynamic viscosity backgrounds. Maximum pressure is 40 pa, and minimum is 0 pa. (b1),(b2) Pressure data charts of the metashell and reference groups along the outer contour of the shell at z = 0.01 cm plane. Red represents the metashell group, and green represents the reference. (c1),(c2) Corresponding velocity distributions. The red arrows represent the normalized velocity. (d1),(d2) Velocity data charts along four lines at x = 0, 3, 6, and 9 cm [shown in (c1)] in the z = 0.01 cm plane. The red lines represent the metashell group. The green lines represent the reference group. Since the reference group has equal velocity everywhere at z = 0.01 cm plane, we only plot one green line for simplicity and clarity. The blue lines at the bottom represent the difference, equal to the velocity of metashell groups minus the velocity of reference groups. }
	\label{f3}
\end{figure}

Next, we validate the performance of the intelligent hydrodynamic concentrator through 3D simulations in real scenarios. See Fig.~\ref{f3}(a1). The entity represents water and the blank represents solid. We alternately arranged the two structures to effectively achieve the dynamic viscosity $\mu_s = {\rm{diag}}(\mu_r^{-1},\mu_\theta^{-1})= {\rm{diag}}(\infty,0)$. For $\mu_\theta^{-1} = 0$, it can be realized by blocking the water flow with solids. For $\mu_r^{-1} = \infty$, it yields $\nabla P = 0$, so we can equivalently realize infinity $\mu_r^{-1}$ by constructing isobaric conditions. We achieve this by raising the height of the water in the metashell to the same level, making it sufficiently higher than the height of the water in the background. In this way, the pressure in the region of the metashell that has the same height as the background is nearly identical, equal to the pressure generated by the water above (Supplemental Material Sec. VI). The straight isobars in Fig.~\ref{f3}(a1) validate the correctness of our design. Similar to Fig.~\ref{f1}(b), we test whether the metashell is still applicable by adding uniformly arranged pillars to the background to effectively change its dynamic viscosity. The result in Fig.~\ref{f3}(a2) shows that the metashell remains functional. To further quantify, we read the pressure data along the outer contour of the metashell and compare it with the reference group. Figs.~\ref{f3}(b1) and \ref{f3}(b2) show the comparison results, where the two sets of data overlap, confirming the accuracy. We also present the corresponding velocity distribution in Figs.~\ref{f3}(c1) and \ref{f3}(c2). The metashell does not affect the velocity in Region III. Similarly, we read the velocity data along four lines at x=0, 3, 6, and 9 cm, and compare the differences with the reference group to quantify the impact on the background. The blue segments at the bottom of Figs.~\ref{f3}(d1) and \ref{f3}(d2) represent the differences, which equal zero in Region III, confirming the correctness of the model. If the metashell is composed entirely of isobaric regions instead of alternating structures, it will disrupt the background physical field (Supplemental Material VI).

Note that the validity of the model is based on the premise that the fluid flow is described by the Hele-Shaw creeping flow model, which typically requires a Reynolds number less than 1. In this work, the calculated Reynolds numbers are all below 0.01, meeting this requirement. Additionally, we achieve isobaric conditions by raising the water height in the metashell region. The raised water height is 0.2 cm, which is significantly smaller than the planar dimensions of the entire Hele-Shaw cell (20 cm $\times$ 30 cm). Therefore, this region can be approximately to satisfy the Hele-Shaw cell height requirement, i.e., $h\ll L$.

Other potential flows that conform to the form of Eq.~(\ref{E1}) are also applicable for designing irregularly shaped intelligent metashell, such as Darcy flow. By introducing pillar arrays to change permeability, experiments have validated hydrodynamic metamaterials designed based on Darcy flow~\cite{ChenPOF24,ChenPNAS22}. Such experimental method can also be utilized to equivalently change the dynamic viscosity, as both changes fundamentally alter fluid transmission capability. Therefore, our model is experimentally feasible. Additionally, besides creating isobaric conditions to equivalently achieve extremely anisotropic dynamic viscosity, other methods, such as using external water pumps~\cite{JiangAM24}, can also achieve the desired effect.

Extremely anisotropic metashells also exhibit transformation-invariant properties~\cite{ZhangPRL19-1}, meaning that the parameter eigenvalues remain unchanged under arbitrary coordinate transformations. This principle allows for designing hydrodynamic metamaterials with other fluid control functions using this metashell, such as rotation~\cite{ParkPRAp19} and guidance~\cite{ChenPOF24,PangPOF24}. Due to the free-form shape of the extremely anisotropic metashell, different fluid control functions can be conveniently achieved by altering the shell's geometry while maintaining the invisibility property. Our model can be extended to multi-physics domain as long as the key parameters conform to transformation theory and effective medium theory. Consequently, it is feasible to design intelligent metashells with irregular shapes for multi-physics coupling, such as thermal-hydrodynamic~\cite{YeungPRAp20,DaiPRAp22,DaiMat23,JinPNAS23} and thermal-electric coupling~\cite{ZhuangIJMSD23,Lei23-1}.

\section{Conclusion}
In summary, this work proposes a scheme for designing hydrodynamic metamaterials using extremely anisotropic dynamic viscosity. This scheme facilitates the design of intelligent hydrodynamic metamaterials and allows for irregular shell shapes, ensuring robustness and interference resistance in complex and variable environments. Additionally, the theoretical requirement for extremely anisotropic dynamic viscosity can be equivalently achieved with a simple method. The proposed model has been validated through finite-element simulations. This approach provides valuable insights for the design of intelligent hydrodynamic metamaterials and multi-physics coupling metamaterials. Furthermore, as discussed earlier, this model is applicable to irrotational creeping flow, offering ideas for designing metamaterials in microfluidic systems and biomedical applications.

\section*{Acknowledgments}

J.H. acknowledges financial support from the National Natural Science 
Foundation of China under Grant Nos. 12035004 and 12320101004 and from the Innovation Program of the Shanghai Municipal Education Commission under Grant No. 2023ZKZD06.
L.X. acknowledges financial support from the National Natural Science Foundation of China 
under Grant Nos. 12375040, 12088101, and U2330401. G.D. acknowledges financial support 
from the National Natural Science Foundation of China under Grant No. 12305046. F. Y. acknowledges financial support from the Postdoctoral Fellowship Program of the China Postdoctoral Science Foundation under Grant No. GZC20242261.
\nocite{*}

\bibliographystyle{apsrev4-1}
\bibliography{ms}
\end{document}